\documentclass[12pt,article,psfig]{aastex}
\newcommand{\kms}{km\,s$^{-1}$}       

\newcommand{\cmthree}{cm$^{-3}$}
\newcommand{\um}{$\mu$m}                                 
\newcommand{\lsun}{$L_{\odot}$}               

\newcommand{\lsim}{\;\lower.6ex\hbox{$\sim$}\kern-7.75pt\raise.65ex\hbox $<$\;}
\newcommand{\gsim}{\;\lower.6ex\hbox{$\sim$}\kern-7.75pt\raise.65ex\hbox $>$\;}
\newcommand{\gapprox}{$\stackrel {>}{_{\sim}}$}   
\newcommand{\lapprox}{$\stackrel {<}{_{\sim}}$}
\newcommand{\asec}{$^{\prime \prime}$}
\newcommand{\adeg}{$^{\circ}$}

\newcommand{\lflux}{W\,cm$^{-2}$}

\newcommand{\lfluxcgs}{ergs\,s$^{-1}$\,cm$^{-2}$}
\newcommand{\fdenscgs}{ergs\,s$^{-1}$\,cm$^{-2}$\,\AA$^{-1}$}
\newcommand{\oia}{[O{\sc i}]63\um}
\newcommand{\oib}{[O{\sc i}]145\um}
\newcommand{\oiii}{[O{\sc iii}]88\um}
\newcommand{\cii}{[C{\sc ii}]158\um}
\newcommand{\sii}{[Si{\sc ii}]34.8\um}
\newcommand{\neii}{[Ne{\sc ii}]12.8\um}
\newcommand{\hii}{H$_2$}

\begin{document}
%
%
%
\shorttitle{FIR Spectroscopy of HH\,1/2} \shortauthors{Molinari
\& Noriega-Crespo}

\title{Far Infrared Spectroscopy of the HH 1/2 Outflow\thanks{ISO is an
ESA project with the instruments funded by ESA  Members States
(especially the PI countries: France, Germany, the Netherlands and the
United Kingdom) and with the participation of ISAS and NASA.}}

\author{Sergio Molinari}
\affil{Istituto di Fisica dello Spazio Interplanetario - CNR, Via Fosso del Cavaliere, I-00133 Roma, Italy}
\author{Alberto Noriega-Crespo}
\affil{SIRTF Science Center \& Infrared Processing and Analysis Center, California Institute of Technology, MS 100-22, Pasadena, CA 91125}

\begin{abstract}

The HH\,1/2 system has been observed with the two spectrometers on board the Infrared Space Observatory. Diffuse
\cii\ emission indicate the presence of a Photo-Dissociation Region (PDR) which is only in part due to the
external Far-UV irradiation from the nearby Orion Nebula. Additional irradiation must be originated locally and
we show that the FUV field produced in the recombination regions behind the shocks traced by the HH objects is
sufficient to induce a PDR at the flow cavity walls.

The analysis of [O{\sc i}], \sii\ and \neii\ lines suggests shock velocities v$_s\sim$100$\div$140 \kms\ with
pre-shock densities between 100 and 1000 \cmthree. H$_2$ pure rotational lines trace a T$\sim$600 K gas which is
likely to be warmed up in slow, planar, molecular shocks. The coexistence in the areas subtended by the
instrumental beamsizes of shocks at different velocities is consistent with the bow-shock morphology of the HH\,1
and, to a lesser extent, HH\,2 as traced by optical images.

\end{abstract}
%

\section{Introduction}
\label{intro}

The HH 1/2 is still the proto-typical optical Herbig-Haro (HH) outflow, and has been and it is the subject of
large number of studies at almost all possible wavelengths, sensitivities and angular resolutions  (see ``Notes''
in Reipurth~\cite{bo99}~for nearly 100 references on this young stellar bipolar flow). Some of the most recent
studies, however, have  concentrated on the molecular gas properties and its relationship to the optical
emission, particularly in the context of entrainment, shock chemistry, proper motions, multiple outflows,
multiple ejection events and the 'unified model' for stellar jets-CO flows. It is in this realm that studies of
the far-infrared emission are of great importance, since in these collisionally excited objects a large fraction
of  the atomic and molecular gas cooling takes place at these wavelengths (see e.~g. Saraceno~\citeyear{sar98};
Molinari et al.~\citeyear{Metal00}).

A quick reminder of the HH 1/2 bipolar jet properties goes as follows; it lies in Orion at $\sim 460$ pc, with an
angular size of $\sim 3\arcmin$ on its brightest component but with signatures of shock excited emission  from a
previous ejection event $\sim 23\arcmin$ NW from VLA 1, the outflow's embedded Class 0 driving source (Ogura
\citeyear{ogu95}; Pravdo et al.~\citeyear{prav85}). The proper motions indicate flow velocities of $\sim 300-400$
\kms\ for both the atomic/ionic and warm H$_2$ gas (Herbig \& Jones~\citeyear{HJ81}; Eisl\"offel, Mundt \&
B\"ohm~\citeyear{eis94}; Rodriguez et al.~\citeyear{rod00}; Noriega-Crespo et al.~\citeyear{nori97}),  and this
sets a dynamical age of $\sim 6\times 10^5$ yrs for the outflow, but with a major ejection event some $\sim
7.5\times 10^4$ yrs ago, defined by the brightest condensations. Recent single-dish radio observations have shown
that  the J=1-0 CO emission is bounded by the ionic/atomic gas  (Moro-Martin et al.~\citeyear{ama99}), thus
providing strong support to the idea that molecular outflows are driven by the stellar jets themselves (see e.g.
Raga~\citeyear{raga94}).

The Infrared Space Observatory (ISO, Kessler et al. \citeyear{Ketal96}) observed HH 1/2 with the mid-IR camera
(CAM) and the short and long spectrometers (SWS \& LWS respectively). A partial analysis of circular variable
filter (CVF) ISOCAM observations  showed the presence of strong ground rotational H$_2$ emission lines with a
morphology quite similar to that of the NIR H$_2$ gas. These data also showed some signatures of strong J-shocks,
like \neii, as expected in high excitation HH outflows (Cernicharo et al.~\citeyear{cerni99}). In this study we
present the analysis of the ISO SWS and LWS data with an emphasis on the properties of shocks in the HH 1/2
outflow and their effects on the immediately surrounding medium. A summary of the observations and data reduction
techniques is presented in $\S$\ref{obs}, followed by a presentation of the results (\S\,\ref{results}) and their
discussion (\S\,\ref{disc}).

\section{Observations and Data Reduction}
\label{obs}

The Long Wavelength Spectrometer (LWS, Clegg et al.~\citeyear{Clegg96}) was used in its LWS01 grating mode to
acquire full low resolution (R $\sim$ 200) 43-197\um\ scans with data collected every 1/4 of a resolution element
(equivalent to $\sim$0.07\um\ for $\lambda$\lapprox 90\um, and to $\sim$0.15\um\ for $\lambda$ \gapprox 90\um); a
total of 25 scans were collected, corresponding to 50s integration time per spectral element.

LWS data processed through Off-Line Processing (OLP) version 7, have been reduced using LIA, the LWS Interactive
Analysis\footnote{LIA is available at http://www.ipac.caltech.edu/iso/lws/lia/lia.html} Version 7.3. The package
consists in a set of tools to inspect in great detail the data quality at the photocurrent level. We have refined
the dark currents and gain for each detector; temporal trends in these two parameters were also estimated and
subtracted/divided. Finally, the data  were recalibrated in wavelength, bandpass and flux. The absolute flux
calibration accuracy for LWS in grating mode is 10\% (ISO Handbook, {\sc iv},  4.3.2).

The Short Wavelength Spectrometer (SWS, de Graauw et al. \citeyear{dGetal96}) was used in the SWS02 grating line
profile observing mode, covering 8 lines: Br$\gamma$, H2 S(1) - S(5), \neii\ and \sii, achieving a spectral
resolution R$\sim 900 - 2000$ (see Table 2.2 ISO Handbook 1.1, p 9). The SWS data was processed using
OSIA\footnote{OSIA is available at http://sws.ster.kuleuven.ac.be/osia}, the SWS Interactive Analysis software,
that permits to revise the performance of the dark current and photometric checks calibration. As in previous
work we have used the latest (November 2000) bandpass calibration files to produce the final spectra.

The final steps of data analysis were done for both LWS and SWS data using ISAP, the ISO  Spectral Analysis
Package Version 2.0\footnote{ISAP is available at http://www.ipac.caltech.edu/iso/isap/isap.html}. Grating scans
(for LWS) and detectors spectra (for SWS) were averaged using a median clipping algorithm optimized to flag and
discard outliers mainly due to transients. Significant memory effects were present in LWS detectors LW2, 3 and 4,
in the form of systematic differences between the two grating scan directions. The LWS observations toward all
positions were heavily fringed, partly due to the presence in the area of the so-called Cohen-Schwartz (C-S) star
(Cohen \& Schwartz \citeyear{CS79}), partly for likely extension of the detected FIR emission; standard techniques
available under ISAP were used to remove these instrumental effects.

\section{Results}
\label{results}

The region observed with the ISO spectrometers is pictured in Fig. \ref{hh1-2}, where the relevant objects are
identified and the locations of the LWS and SWS beams are also indicated. The C-S star is a $\sim$30 \lsun\
T\,Tauri star (Cohen et al. \citeyear{Cetal84}) that was believed to be the exciting source of the bipolar flow
before VLA\,1 was discovered by Pravdo et al. (\citeyear{prav85}).

\placefigure{hh1-2}

Line detection was established by visual inspection of the averaged spectra, and Gaussian functions were fitted
to estimate integrated line fluxes. In order for a line to be considered as detected, it had to be consistently
present in both scan directions. This is particularly critical for lines in the LWS spectral range between
$\sim$110 and 160\um, where detector memory effects were clearly identified (\S\ref{obs}). Figure \ref{lines}
show the lines detected with the two spectrometers.

\placefigure{lines}

To estimate uncertainties we multiplied the standard deviation of the fit from the data by the width of the
spectral resolution element. Computed values are reported in Table \ref{restab}. Line and continuum flux was
observed from all LWS pointed positions. The LWS beam has a nearly Gaussian shape, so that radiation may be
collected from outside the nominal FWHM beamsize. At the distance of one beam this contamination will be of the
order of 10-20\%, and since the observed positions with the LWS are adjacent, we can in principle estimate the
true flux coming from within the individual beams. In the case of the HH\,1 and HH\,2 pointings, contamination is
only due to the flux coming from VLA\,1 (the two HH objects are too far to reciprocally contaminate each other).
Instead, the observed flux from VLA\,1 will be contaminated from both HH\,1 and HH\,2. The contamination
corrected flux for the lines detected with the LWS are reported in italic in Table \ref{restab}

\placetable{restab}

\cite{GNL01} have recently independently reduced the same ISO-LWS data toward HH\,1/2. Compared to theirs, our
reduction includes the full data reprocessing through the LWS Interactive Analysis (\S\,\ref{obs}) which is
crucial to really have a handle on the level of reliability of the calibrated data. By averaging scan directions
separately we were able to estimate the line fluxes using the more reliable scan direction when detector memory
effects were identified (\S\ref{obs}). Maybe as a result of this higher level of analysis, we have not reported
as detected in Table \ref{restab} a number of lines (mainly CO lines) which are instead reported as detected by
\cite{GNL01}. However, for the lines commonly reported as detected the fluxes are in substantial agreement
(within 10\%) for the brighter lines; for fainter lines the discrepancy can be as high as a factor two, but the
relative uncertainties in the flux estimates are also as large.

Few basic considerations will help set the frame for the subsequent discussion. \cii\ is detected toward all LWS
pointings. Since \cii\ line is one of the main coolants in FarUV-irradiated gas, it seems natural to postulate
its origin in a Photo-Dissociation Region (PDR, see e.g. Tielens \& Hollenbach \citeyear{TH85}). Post-shock
cooling regions may also produce \cii, but this is not relevant in the present case because the implied levels of
\oia, expected to be at least ten times stronger than \cii\ for a wide range of shock velocities and pre-shock
densities (e.g., Hollenbach \& McKee \citeyear{HM89}), are simply not observed. The origin of [O{\sc i}] lines,
also detected toward all observed positions, is more difficult to assess; \oia, definitely the main shock coolant,
can be an important (and even dominant) cooling line in PDRs too, if the density is n\gapprox $10^4$ \cmthree\ and
the FUV irradiation level, expressed in units of 1.6$\;10^{-3}$\lfluxcgs, the local interstellar radiation field
(Habing \citeyear{H68}), is G$_0$\gapprox $10^2$ (Kaufman et al. \citeyear{Ketal99}). There is very little doubt,
however, that a major contribution to [O{\sc i}] emission comes from shocks since \sii\ and \neii, both important
coolants in dissociative shocks, are also detected in the present observations; a conclusion further supported by
the theoretical J-shock modeling (Hartigan, Raymond \& Hartmann~\citeyear{HRH87}; Hartigan, Morse \& Raymond
\citeyear{HMR94}) applied to the strong optical [O{\sc i}] 6300+6363 ~\AA\ emission along the entire outflow (Solf
\& B\"ohm \citeyear{SB91}). So to explain the observed FIR line emission we need to invoke at least two physical
mechanisms; a PDR for\cii\, and shocks for at least part of [O{\sc i}]. A detailed analysis of the physical
conditions of the PDRs in which the \cii\ line is generated will provide an estimate of the relative fractions of
[O{\sc i}] lines radiated by the two mechanisms.


\section{Discussion}
\label{disc}
\subsection{The PDR}
\label{pdr}

\cii\ line has similar fluxes, within 20\%, along the flow; in particular, the line is stronger toward HH\,2 and
fainter toward VLA\,1. On this basis, we exclude both the C-S star and the heavily embedded VLA\,1 source as the
FUV (between 6eV and 13.6eV) field sources responsible for the PDR conditions. Either the FUV field source is
external to the HH\,1/2 flow, or, if local, must be somehow distributed along the flow. We will see below that a
combination of the two possibilities best explains the observations.

\subsubsection{External irradiation}
\label{pdr-external}

The external irradiation hypothesis is supported by the evidence (Mundt \& Witt \citeyear{MW83}) of a significant
UV irradiation on the HH\,1/2 area from the Orion nebula, whose center lies $\sim1.5$\adeg\ to the north. This
contamination would amount to 30$\div$50\% of the flux levels detected toward HH\,1/2 shortward of $\sim$1500\AA,
decreasing to $\sim$20\% at longer wavelengths. Based on the IUE measured spectra (B\"ohm-Vitense et al.
\citeyear{Betal82}; B\"ohm, Noriega-Crespo \& Solf~\citeyear{BNS93}) the integrated FUV field due to irradiation
from the Orion Nebula would amount to G$_0\sim$7, although it should be noted that the data used by \cite{MW83}
do not go shortward of $\sim 1500$\AA. This value is compatible with the FUV field from the dominant Trapezium
star ${\rm \theta}_c^1$ (about G$_0\sim18$), with a reasonable amount of extinction along the path to HH\,1/2. The
maximum \cii\ line flux in the ISO-LWS beam (assuming complete beam filling) that can be produced by this FUV
field is $\sim3.5\;10^{-19}$ \lflux\ for a gas density n$\sim10^3$ \cmthree\ (Kaufman et al. \citeyear{Ketal99}).
The \cii\ flux decreases by about 40\% for densities a factor 10 higher or smaller, and is almost linear with
G$_0$. We conclude that about half, at most, of the PDR \cii\ emission measured by ISO-LWS may be due to external
irradiation by the Orion Nebula.

\subsubsection{Local irradiation}
\label{pdr-local}

In a recent paper (Molinari, Noriega-Crespo \& Spinoglio \citeyear{MNS01}) we proposed that the \cii\ flux
detected toward the HH objects in the HH\,80/81 flow was emitted in a PDR situated at the walls of the cavity
excavated by the flow, and irradiated by collisionally-enhanced 2-photon FUV continuum emitted by the post-shock
ionized regions in the HH objects (Dopita, Binette \& Schwartz \citeyear{DBS82}). We propose that this scenario
is also pertinent to the HH\,1/2 system. In terms of observables, this model establishes a relationship between
the free-free radio continuum emitted by the recombination region in the HH object, and the \cii\ line flux
radiated by the PDR at the cavity walls. In the optically thin regime, the former can be expressed as (Curiel,
Cant\'o and Rodr\'{\i}guez \citeyear{CCR87}):

\begin{equation}
S_{\nu} = 1.84\cdot 10^{-4} ~\theta^2~ \left[{{\nu}\over {10~ GHz}}\right]^{-0.1}~ T_4^{0.45}~ n_{o_{10}}
~v_{s_7}~ [1+3.483 v_{s_7} -2.745]~{\rm mJy} \label{radioshock}
\end{equation}

$\theta$ is the angular diameter of the recombination region and is estimated off-the-plot from the 6 cm VLA
D-conf. maps of \cite{Retal90} to be $\sim$10\asec. T$_4$ is the electrons temperature in units of 10$^4$ K and we
hold it fixed at 1 in these units. $v_{s_7}$ is the shock velocity in units of 100 \kms\ and we fix it to 1 in
these units as determined in \S\,\ref{atomic} below. $n_{o_{10}}$ is the pre-shock density in units of 10
\cmthree\ and we will keep this as the only free parameter in Eq.(\ref{radioshock}).

The \cii\ line flux can be expressed as (Molinari, Noriega-Crespo \& Spinoglio \citeyear{MNS01}):

\begin{equation}
F_{C{\sc ii}} = {{10^{-7}}\over {D^2}} {{4\pi}\over{3}} r_{HH}^3 \; \eta_{ce} \; f_c \;\chi_c \int_{6eV}^{13.6eV}
j_{\nu} d\nu~{\rm W\,cm^{-2}}
 \label{fcii}
\end{equation}

D is the distance of the source from the Sun. j$_{\nu}$ is the 2-photon process emissivity and is a function,
among other things, of the square of gas density. f$_c$ is the fraction of the 2-photon continuum emitted by the
HH object which is intercepted by the portion of flow cavity wall encompassed by the instrumental beam used to
measure the \cii\ flux, and can be derived with simple geometrical considerations. The radius of the HH object is
set to 5\asec\ (see above).

A critical parameter in Eq.(\ref{fcii}) is the factor $\eta_{ce}$ which accounts for the collisional enhancement
of the hydrogenic 2s level population above the values predicted by pure recombination. $\eta_{ce}$ can be
determined from a comparison of the predicted two-photon spectrum with the observed UV continuum from HH objects.
\cite{DBS82} find values $5.5\leq\eta_{ce}\leq10.6$ for the different knots of HH\,1, and $2.8 \leq \eta_{ce}
\leq6.2$ for the different knots of HH\,2. Since most of the knots in the two pointings will be encompassed by
the LWS beam, we will adopt the mean $\eta_{ce}$ values.

Another important parameter is the fraction $\chi_c$ of the incident FUV field which is re-radiated via the \cii\
line; this parameter is predicted to vary between 0.1\% and 1\% (Tielens \& Hollenbach \citeyear{TH85}). Since
the rest of the FUV flux is reprocessed by dust into Far-IR continuum, our data should in principle allow us to
independently estimate $\chi_c$ as the ratio between the \cii\ line flux and the integrated FIR continuum as
revealed by the LWS. Indeed, LWS observations clearly reveal significant continuum emission longward of
$\sim$50\um\ toward all pointings. The full spectra are shown in Fig. \ref{lws_continua_wcm2um}, where the
minispectra from the ten detectors have been stitched together for cosmetic purposes. The region is quite complex
at these wavelengths and our continuum data do not represent an improvement, given the relatively poor spatial
resolution, with respect to previous studies. Our data are in substantial agreement with \cite{Cetal84}, also
considering that our pointed positions do not match the emission peaks of the bipolar emission pattern mapped at
100\um\ with the KAO. The peak wavelengths in the SEDs of Fig. \ref{lws_continua_wcm2um} imply black-body dust
temperatures of 25, 32 and 47 K toward HH\,2, VLA\,1 and HH\,2 respectively, in agreement with the KAO
observations of \cite{Hetal86}. The origin of the FIR continuum is not easy to assess. \cite{Cetal84} do not
detect 100\um\ emission from HH\,2 to a 3$\sigma$ limit of $\sim$12Jy; clearly, a portion of the continuum that
we detect toward HH\,2 is due to contamination from VLA\,1. As concerns HH\,1 it is clear form the KAO data that
a significant portion of the FIR continuum comes from a radio source $\sim$30\asec\ S-SW of HH\,1 and associated
with an H$_2$O maser (Pravdo et al. \citeyear{prav85}). Based on the same dataset, it seems also clear that the
C-S star does not seem to significantly contribute longward of 50\um; this is not surprising given its state of
relatively evolved PMS star.

\placefigure{lws_continua_wcm2um}

The pointing where a more reliable estimate of the PDR FIR continuum can be derived is probably HH\,2. Indeed, the
HH\,1 continuum is severely contaminated by two sources (see above), while the VLA\,1 continuum is likely
dominated by thermal dust emission in the envelope of this Class\,0 source. HH\,2 instead, may only partially be
contaminated by the VLA\,1 continuum. The \cii/FIR ratio toward HH\,2 yields $\chi_c\geq0.005$, clearly in the
range of values expected from the PDR models.

With this choice of parameters we run our models (Molinari, Noriega-Crespo \& Spinoglio \citeyear{MNS01}) to
determine the quantities in Eqs.(\ref{radioshock}) and (\ref{fcii}) as a function of the gas density n (in
\cmthree) and shock compression ratio $\mathcal{R}$=n/n$_0$. The resulting grids are presented in Fig.
\ref{ciiradiofig} (full-line for HH\,2, dashed-line for HH\,1). The figure also shows the location of the HH
objects based on the 6 cm flux observed by \cite{Retal90}, and the \cii\ observed with the LWS, decreased by an
amount $\sim3.5\;10^{-19}$ \lflux\ which is our estimate of the maximum contribution due to external FUV
irradiation (\S\ref{pdr-external}).

\placefigure{ciiradiofig}

The adopted model seems successful in reproducing the observable quantities. Formal values of density and
compression ratio are [n, $\mathcal{R}$]=[(5200,5100), (40,18)] for HH\,1 and 2 respectively. A problem with this
scenario, however, is that we seem to produce too much FUV continuum. The predicted \emph{emitted} flux at the
indicative wavelength of 1500\AA\ for the adopted model parameters is $\sim 8\,10^{-13}$ \fdenscgs, which should
be compared with the dereddened observed values of $\sim10^{-13}$ \fdenscgs\ for the two HH objects (e.g.,
B\"ohm-Vitense et al. \citeyear{Betal82}). We can tune the model parameters to produce less FUV continuum and
still fit the \cii\ and 6 cm observations. For example, we may decrease the radii of the ionized regions by 50\%,
which is not unreasonable since we estimated this parameter directly from plots of radio maps. Additionally, we
can also increase $\chi_c$; in low irradiation conditions, like in the present case, this parameter can reach
values $\sim0.01$ (Tielens \& Hollenbach \citeyear{TH85}) which is not in contradiction with our previous
estimate as this was a lower limit due to the difficulty of assessing the intrinsic FIR continuum of the HH
objects (see above). With this choice of parameters we still fit the observed \cii\ line flux and 6 cm radio
continuum, but with densities a factor 2 higher and compression ratios a factor 2 lower, producing an emitted
1500\AA\ continuum of $\sim3\,10^{-13}$ \fdenscgs; the discrepancy with UV observations is reduced to a factor 3,
which could be partly accounted for by the uncertainties in the reddenening corrections applied.

As concerns the jet emanating from VLA\,1, its low ionization (Solf \& B\"ohm \citeyear{SB91}) also confirmed by
the low radio flux measured (e.g., Rodr{\'{\i}}guez et al. \citeyear{Retal90}) is not sufficient to produce the
amount of FUV continuum that would be required by our models to induce the \cii\ emission observed toward VLA\,1.
In the absence of significant dust column along the line of sight from the HH objects to the proximity of VLA\,1,
however, the direct FUV irradiation from the HH objects onto the cavity walls at the base of the flow would be
sufficient to justify the \cii\ line flux observed toward VLA\,1. The observation of a scattered-light component
in the [S{\sc ii}]$\lambda\lambda$6716,6731 line along the jet (Solf \& B\"ohm \citeyear{SB91}), and the presence
of a biconical nebula seen in scattered optical light at the base of the jet (Strom et al. \citeyear{Setal85})
would indeed seem to strengthen the idea of a relatively dust-free flow cavity (see also Davis, Eisl\"offel \&
Ray \citeyear{DER94}).

\subsubsection{Line cooling}
\label{pdr-lines}

We are now able to estimate the amount of cooling expected in lines other than \cii\ in the PDR conditions
discussed above. The external and relatively faint FUV field (G$_0\sim 7$) from the Orion Nebula can generate
F(\oia)$\sim$5\%F(\cii), at most, for n$\sim10^3$ \cmthree. Increasing the density will also increase the
expected \oia/\cii\ cooling ratio (Kaufman et al. \citeyear{Ketal99}), but the absolute \cii\ flux emitted would
also decrease by a similar amount living the absolute \oia\ cooling essentially unaltered. As concerns the local
FUV field, the G$_0$ value for the PDR will be dependent on the assumed distance between the HH objects and the
flow cavity walls. The CO maps of \cite{ama99} reveal a full width of the low-velocity lobes in the HH\,1/2
outflow of the order of $\sim30$\asec, implying G$_0\sim40$ at the cavity walls and a maximum \oia\ PDR cooling
$\sim20\div30$\% of the \cii. As noted above, the \emph{absolute} [O{\sc i}] cooling is almost independent on the
assumed density.

Finally, high-J CO and rotational H$_2$ lines fluxes comparable to the observed values can only be generated in
energetic (G$_0\geq10^3$) and very dense (n$\geq10^6$ \cmthree) PDRs (e.g., Burton, Hollenbach \& Tielens
\citeyear{BHT90}, which is clearly not the present case.

\subsection{Shock conditions in the HH objects}
\label{shock}

\subsubsection{Atomic lines}
\label{atomic}

The shock conditions toward HH\,1 and 2 were diagnosed comparing our observations (after correcting the [O{\sc i}]
lines for PDR contamination, see \S\ref{pdr-lines}) with the predictions from plane parallel atomic shock models.
The shock models were calculated using MAPPINGS2, a code developed by \cite{BDT85}, which has been  thoroughly
tested (Pequignot \citeyear{P86}). A grid of models was created changing the pre-shock density between 10$^2$ and
10$^4$ \cmthree, and the shock velocity between 60 and 140 \kms. The rest of the model parameters have been fixed
to standard values (Molinari, Noriega-Crespo \& Spinoglio \citeyear{MNS01}). The \oia/\oib$-$\neii/\sii\ diagram
in Fig. \ref{hh1-2_diag} shows that shock velocities of v$_s\sim$100 or $\sim$140 \kms, and pre-shock densities
between 10$^2$ and 10$^3$\cmthree\ would be adequate to describe the observed line ratios. Higher shock
velocities seem to be ruled out since \neii/\sii\ ratios largely in excess with respect to the observations are
produced (e.g., $\sim$10 for v$_s\sim$220 \kms, using the planar shock models of \cite{HRH87}). We note that the
beamsize of the SWS at the \sii\ wavelength is a factor $\sim1.7$ larger than at the \neii\ wavelength, but this
would not change the position of the HH objects in Fig. \ref{hh1-2_diag} by a significative amount even for
uniformly extended emission. Noticeably, the estimated pre-shock densities are in good agreement with the
independent estimates based on the shock-excited PDR model (\S\ref{pdr-local}). Those models were computed using
v$_s$=100 \kms, but we note that the adoption of v$_s$=140 \kms\ would simply imply, according to Eq.
(\ref{radioshock}), a right-shift of the model grids in Fig. \ref{ciiradiofig}, still fitting the observations
(with higher compression ratios, though).

\placefigure{hh1-2_diag}

The comparison with optical line diagnostics is not straightforward given the complex morphology of the HH objects
which appear structured in several knots, both in optical and near-IR images, which our observations do not
spatially resolve (Herbig \& Jones \citeyear{HJ81}; Noriega-Crespo \& Garnavich \citeyear{NG94}; Davis,
Eisl\"offel \& Ray \citeyear{DER94}). The SWS pointing toward HH\,1 encompasses knots A, C, D and F; optical
lines suggest shock velocities $\sim80$\kms\ for A, and consistently above 130\kms\ for all the others (Hartigan,
Raymond \& Hartmann \citeyear{HRH87}). The dust scattered emission requires at least 150 \kms\ shocks to explain
the optical observations (Noriega-Crespo, Calvet \& B\"ohm \citeyear{NCB91}). Our estimate of v$_s\sim100$\kms\
might represent average physical conditions over the SWS beamsize, although the negative detection of \oiii\ in
the present observations (expected to be at the same flux level of the \oia\ line) cannot in principle be
explained with v$_s\geq$130\kms. The presence of strong [O~III] 4959+5007 \AA, however, in HH 1F and HH 2A+H
(Solf \& B\"ohm \citeyear{SB91}) sets a lower limit at $\sim 100$~\kms\ for the shock velocity and suggests that
other mechanisms like, e.g., partial ionization of the pre-shock material, may be responsible for the depression
of \oiii\ flux levels. Likewise, the SWS pointing toward HH\,2 includes knots A, B, C, G, H and M, where shock
velocities estimated from optical lines range from 100 to 200 \kms~(Hartigan, Raymond \& Hartmann
\citeyear{HRH87}; Solf \& B\"ohm \citeyear{SB91}).

\subsubsection{Molecular lines}
\label{molecular}

Pure rotational H$_2$ lines in the ground vibrational level are normally analyzed under LTE conditions (e.g.,
Gredel~\citeyear{G94}). Einstein coefficients and wavenumbers were taken from \cite{BD76} and \cite{D84}, and an
ortho/para=3 has been assumed. The SWS aperture at the S(1) and S(2) line wavelengths is a factor $\sim1.7$
larger compared with the other H$_2$ lines; fluxes for these two lines were corrected accordingly, assuming
extended and uniform emission. We dereddened the H$_2$ line fluxes using the visual extinctions given by
\cite{Betal87} and using the \cite{RL85} extinction curve. The H$_2$ gas temperature and column density can then
be directly obtained from the slope and intercept of a linear fit to the Boltzmann plots shown in
Fig.~\ref{h2_excitation}, and are reported in Table \ref{h2physparam} together with 1$\sigma$ uncertainties in
parenthesis; a solid angle equal to the SWS focal plane aperture for the S(3)-S(5) lines was assumed,  i.e.
14\asec\ x 20\asec~or 6.6$\times10^{-9}$ sr (valid for all detected  \hii\ lines but S(1), where the solid angle
is $\sim$30\% higher), with a beam filling  factor of 1. The fits show that a single temperature component is
adequate to represent the observations, given the observational uncertainties.

\placefigure{h2_excitation}

Using the same model and the parameters determined from the Boltzmann plots, we can estimate an H$_2$ cooling, for
the component traced by the (0-0) lines, of $9\;10^{-3}$ \lsun\ and $2.5\;10^{-2}$ \lsun\ for HH\,1 and HH\,2,
respectively.

\placetable{h2physparam}

The issue of the nature of the H$_2$ excitation in HH\,1 and 2 has been discussed by several authors in the last
decade using near-IR imaging of the S(1)1-0 2.12\um\ and near-IR spectra in the H and K bands (Noriega-Crespo \&
Garnavich \citeyear{NG94}; Davis, Eisl\"offel \& Ray \citeyear{DER94}; Gredel \citeyear{G96}; Eisl\"offel, Smith
\& Davis \citeyear{ESD00}; Davis, Smith \& Eisl\"offel \citeyear{DSE00}). The spectra suggest excitation
temperatures ranging between 2000 and 3000 K in the various emission knots revealed in the images. Detailed
modeling suggests that C-type (Draine \citeyear{D80}) shocks best fit the observational scenario; H$_2$ emission
would originate in the tails of the bow-shocks, whose tips are instead bright in [Fe{\sc ii}]1.64\um\ (Davis,
Eisl\"offel \& Ray \citeyear{DER94}), where the shock velocities are low enough for H$_2$ not to be dissociated.
The fact that Boltzmann plots based on H$_2$ ro-vibrational are satisfactorily fitted with single linear fits led
\cite{G96} to doubt as to the C shock origin, since a stratification of temperatures up to $\sim3000$ K should be
expected. \cite{ESD00} review the most popular shock models and indeed confirm that curved C-type shocks produce a
quite shallow trend in the Boltzmann plots; besides, the predicted excitation temperatures of the $v=1$ lines are
significantly lower ($\sim1000$ K) than actually observed. Planar C-type shocks, on the other hand, seem to
produce a rather abrupt change of slope at upper energy levels $\sim4000-5000$ K. Since this is exactly at the
separation between the $v=0$ lines we detected in the present work and the $v>0$ lines detected in the near-IR,
the combined analysis of mid- and near-IR H$_2$ lines seems to suggest an origin in planar C-type shocks.

C-type shocks are good emitters of molecular lines in general. Our marginal detection of high-J CO lines at flux
levels comparable to H$_2$ (0-0) lines suggests shock velocities $15\leq v_s \leq 20$ \kms\ and pre-shock
densities n$_0\geq 10^5$ \cmthree\ (Kaufman \& Neufeld \citeyear{KN96}). However, for the same parameters the
model also predict an H$_2$O(2$_{12}$-1$_{01}$) line flux about two orders of magnitude higher than actually
observed in HH\,2 (Tab. \ref{restab}). Unless H$_2$O vapor is confined in a much smaller area compared to H$_2$
or CO, there is a clear deficit with respect to expectations. This seems unusual for Class-0 objects (Giannini,
Nisini \& Lorenzetti \citeyear{GNL01}) and resembles more evolved Class-I systems. The possibility that water
vapor depletes onto grains in the post-shock cooling region has been verified for HH\,7 in NGC\,1333 with the
detection of the 62\um\ ice band (Molinari et al. \citeyear{Metal99}); we do not detect such a feature anywhere
in the HH\,1/2 system, although the filling factor of the bow-shocks present in HH\,1 and 2 (e.g., Fig.
\ref{hh1-2}) in the LWS beam is lower than in the case of HH\,7 (also because of the larger distance). An
interesting alternative to explain the water vapor deficit could be offered by the marginal detection of the
fundamental lines of the OH molecule in HH\,2. Given the high velocity of the dissociative shocks in HH\,1/2,
significant amounts of Far-UV flux can be expected to be irradiated by the post-shock regions (see
\S\,\ref{pdr}); in particular, the photons between 1300\AA\ and 2000\AA\ are capable of dissociating the water
molecule into H and OH (Andresen, Thissen \& Schroeder \citeyear{ATS01}). Water photodissociation by a UV field
was also invoked by \cite{Setal00} to explain the surprisingly high OH cooling found toward T\,Tau.

Comparing the $v_s\sim$20 \kms\ found for the C-type molecular shocks with the $v_s\sim$100 \kms\ deduced for the
dissociative J-type shock traced by the atomic lines (\S\,\ref{atomic}), it is plausible that the two shock
regions are respectively associated with the tails and tips of bow-shocks. Such structures are indeed identified
in HH\,1 and, with a more complex morphology, in HH\,2 (Hester, Stapelfeldt \& Scowen \citeyear{HSS98}). The 2
orders of magnitude difference in pre-shock densities implies that significant compression of the pre-shock
material has taken place in the molecular shocks. This is expected given the very nature of the C-type shocks,
where the relatively low shock velocity and the presence of transverse magnetic field combine to accelerate and
pre-compress the upstream material. I-band polarimetry (Strom et al. \citeyear{Setal85}) indeed confirms the
presence of a magnetic field oriented along the HH\,1/2 flow.

\acknowledgments{S.M. thanks the generous support of NASA Jet Propulsion Laboratory through contract 1227169.
A.N-C. research has been supported in part by NASA ADP grant NRA-00-01-ADP-096. The paper's clarity considerably
improved thanks to the comments of an anonymous referee. ISAP is a joint development by the LWS and SWS
Instrument Teams and Data Centers. Contributing institutes are CESR, IAS, IPAC, MPE, RAL  and SRON. LIA is a
joint development of the LWS consortium. Contributing institutes are RAL, IPAC and CESR. OSIA is a joint
development of the SWS consortium. Contributing institutes are SRON, MPE and KUL.}

\clearpage

%
%

\figcaption[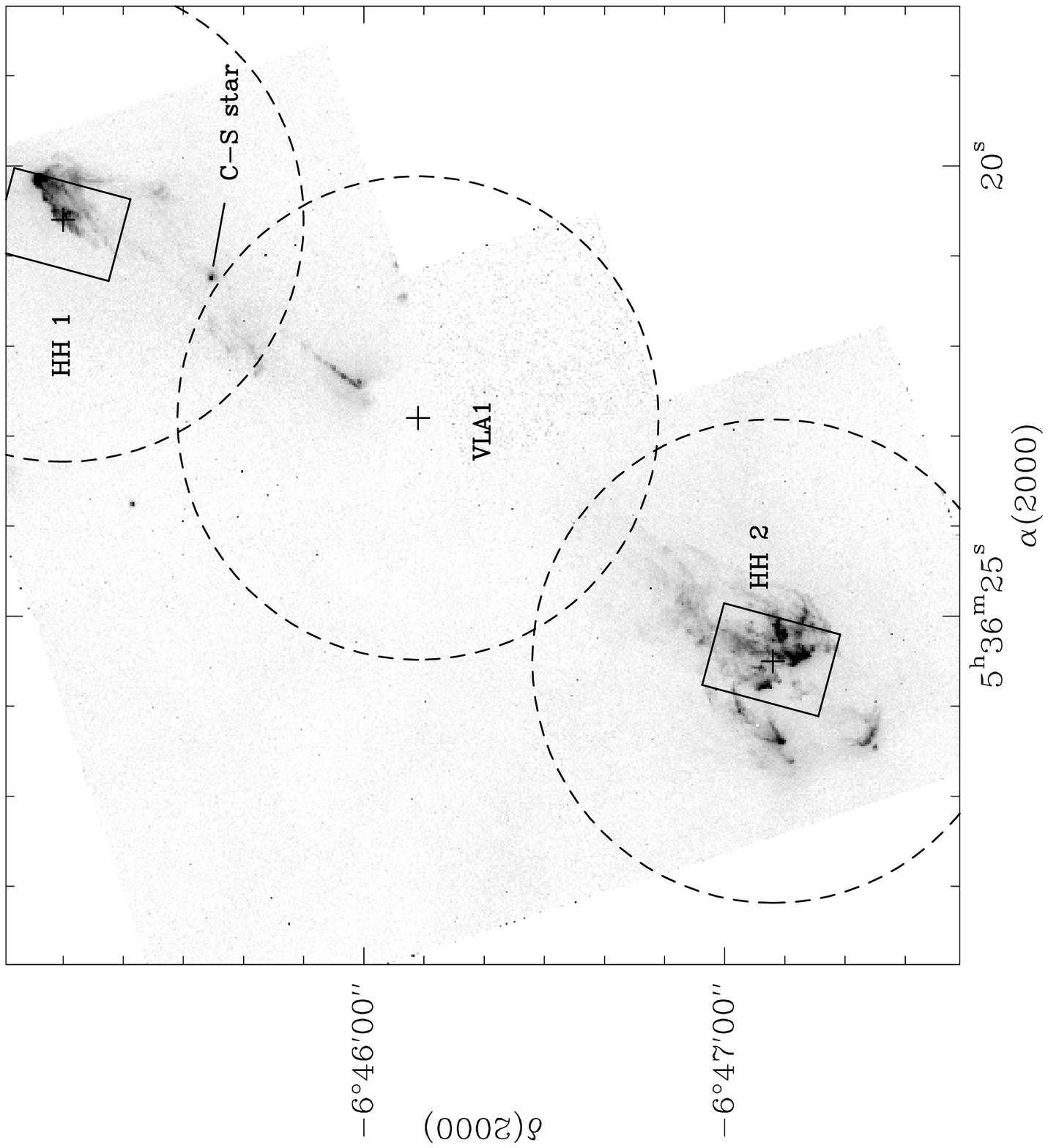]{\label{hh1-2} HST [S{\sc ii}] image of the HH\,1/2 region. The crosses mark the
pointed positions, while dashed circles and full-line rectangles indicate the FWHM beamsizes of the LWS and SWS
respectively. The position of the C-S star is also indicated.}

\figcaption[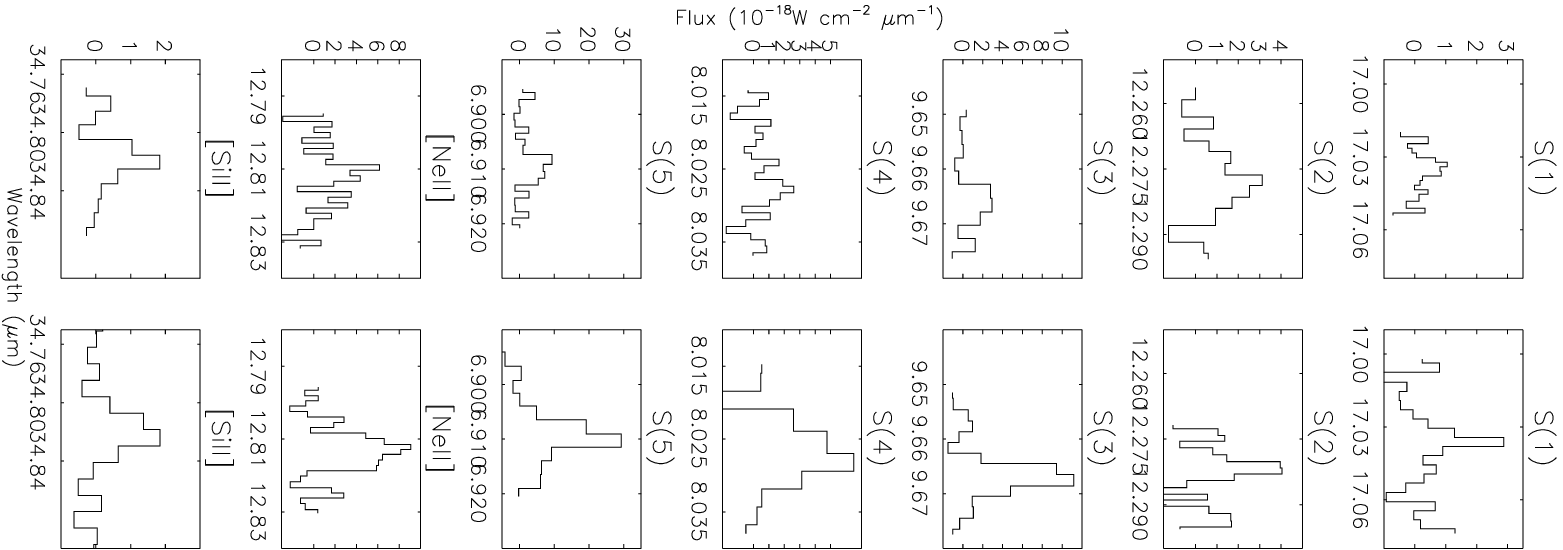]{\label{lines} {\bf a.} H$_2$, \neii\ and \sii\ lines detected toward HH\,1 (left
column) and HH\,2 (right column).}

\addtocounter{figure}{-1} \figcaption[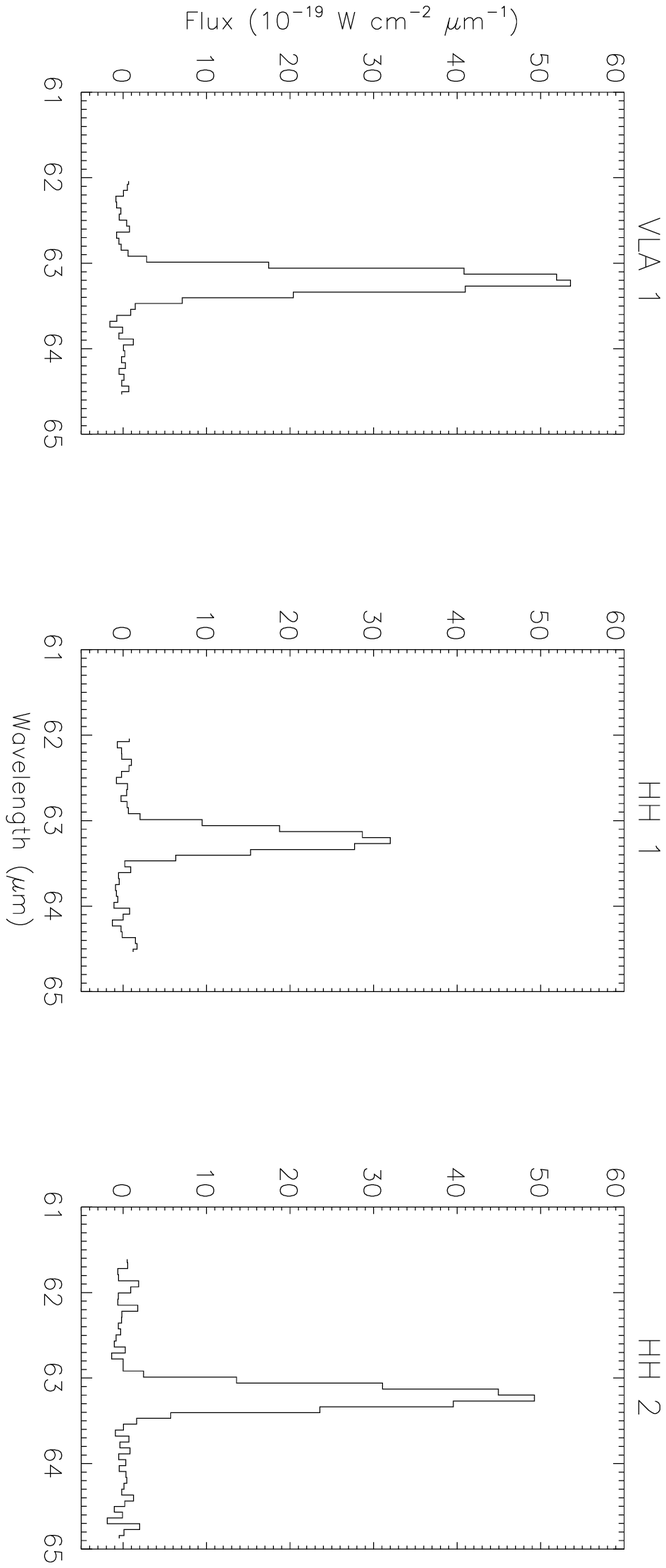]{{\bf b.} \oia\ lines detected toward the three positions
observed with the LWS.}

\addtocounter{figure}{-1} \figcaption[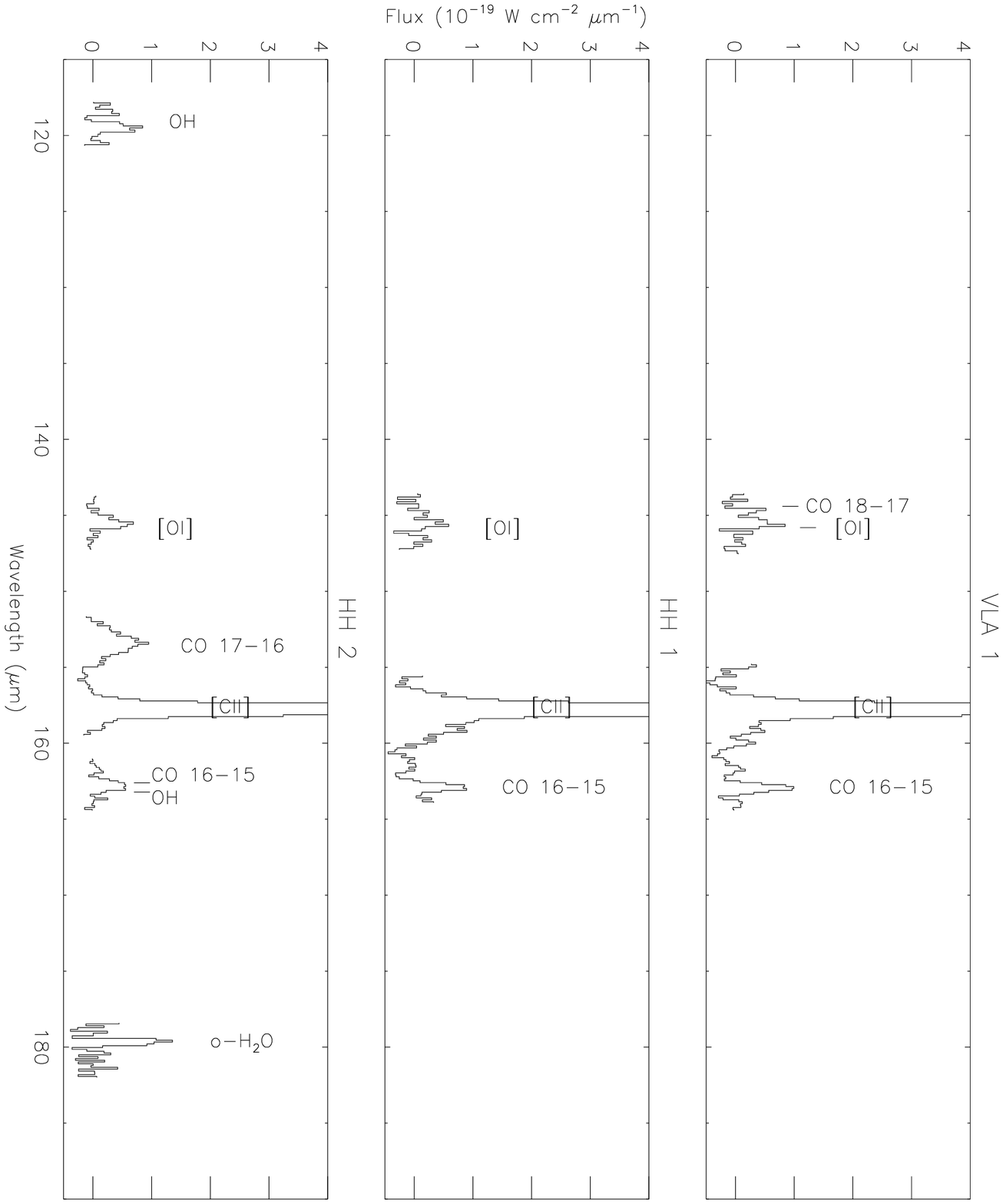]{{\bf c.} Lines detected between 110 and 180\um\ toward
the three positions observed with the LWS.}

\figcaption[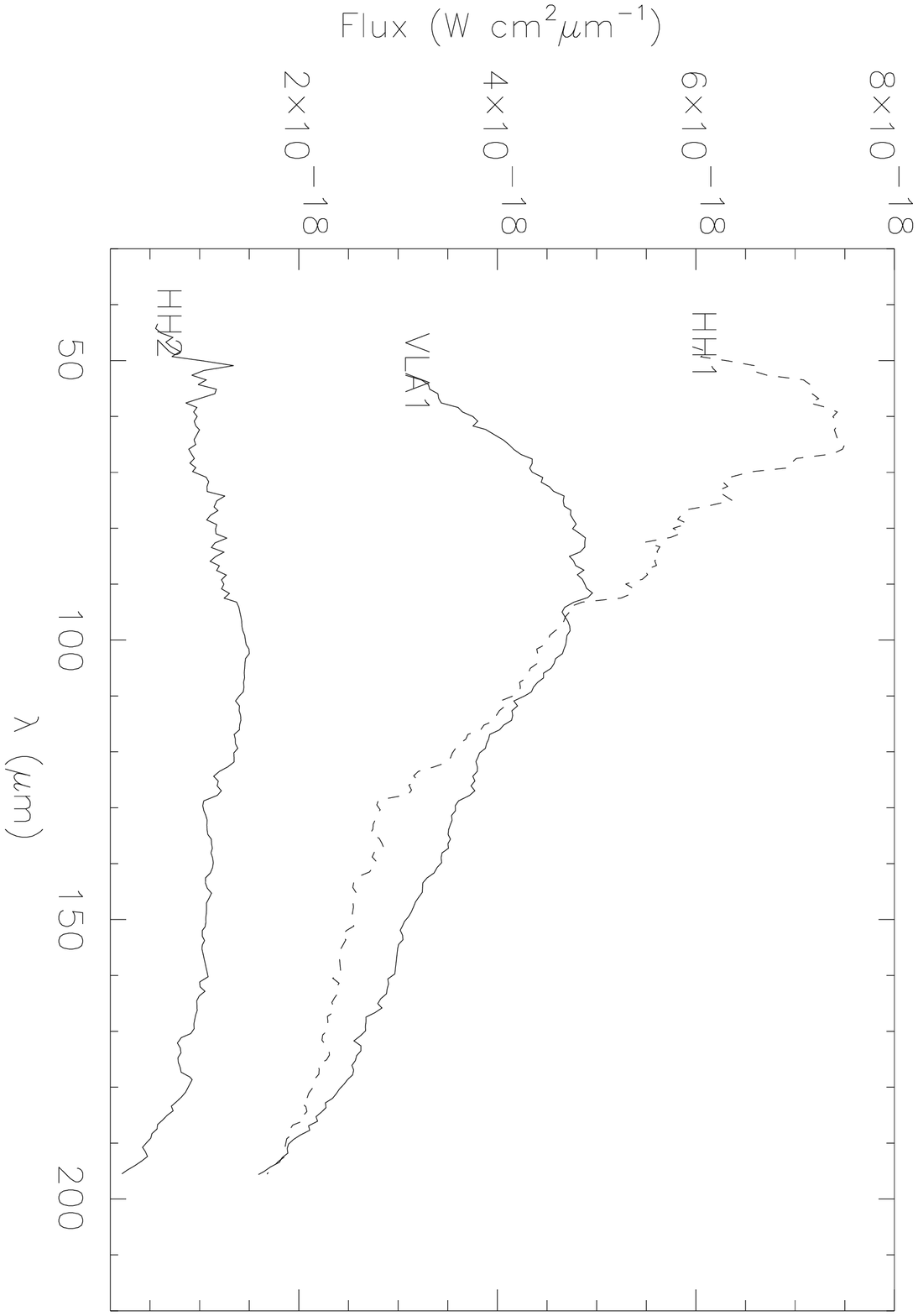]{\label{lws_continua_wcm2um} Spectral energy distributions observed with LWS toward
all pointings. The 10 LWS detectors have been stitched together to produce a smooth SED. No correction has been
applied for the reciprocal contamination between adjacent beams.}

\figcaption[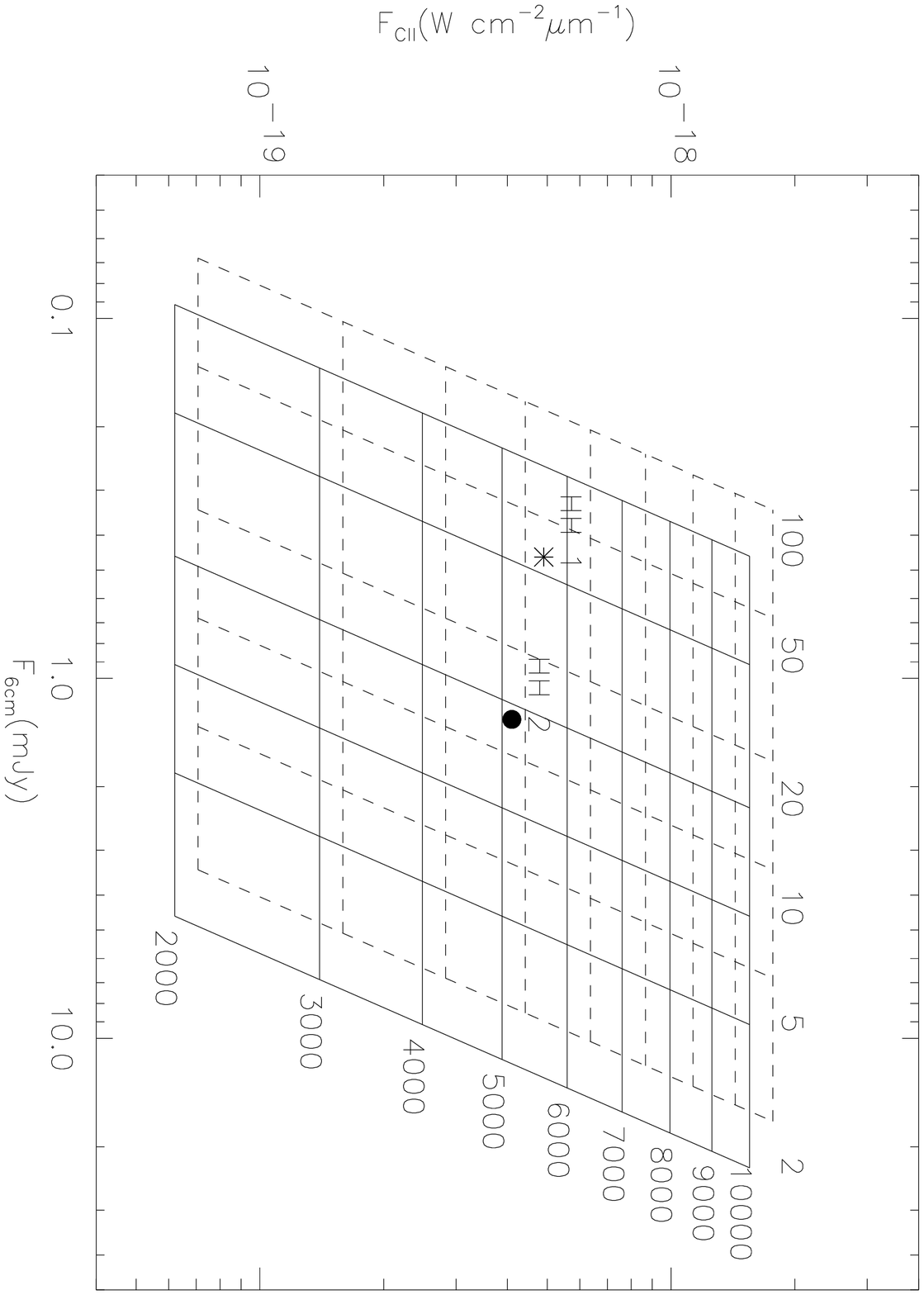]{\label{ciiradiofig} \cii\ line flux as a function of 6 cm radio continuum flux
expected for PDR emission induced by the FUV two-photon continuum from the ionized regions in the HH objects. The
grids show the loci of points of constant density $n$ (horizontal lines) and constant $\mathcal{R}$=n/n$_0$
compression ratio in case of a recombination region 5\asec\ in radius behind a shock front in a HH object with
v$_s$=100 \kms. The other parameters used are [$\eta_{ce}$, $\chi_c$]=[(8.0,4.5),(0.005,0.005)] for the dashed
and full-line grids (HH\,1 and HH\,2) respectively.}

\figcaption[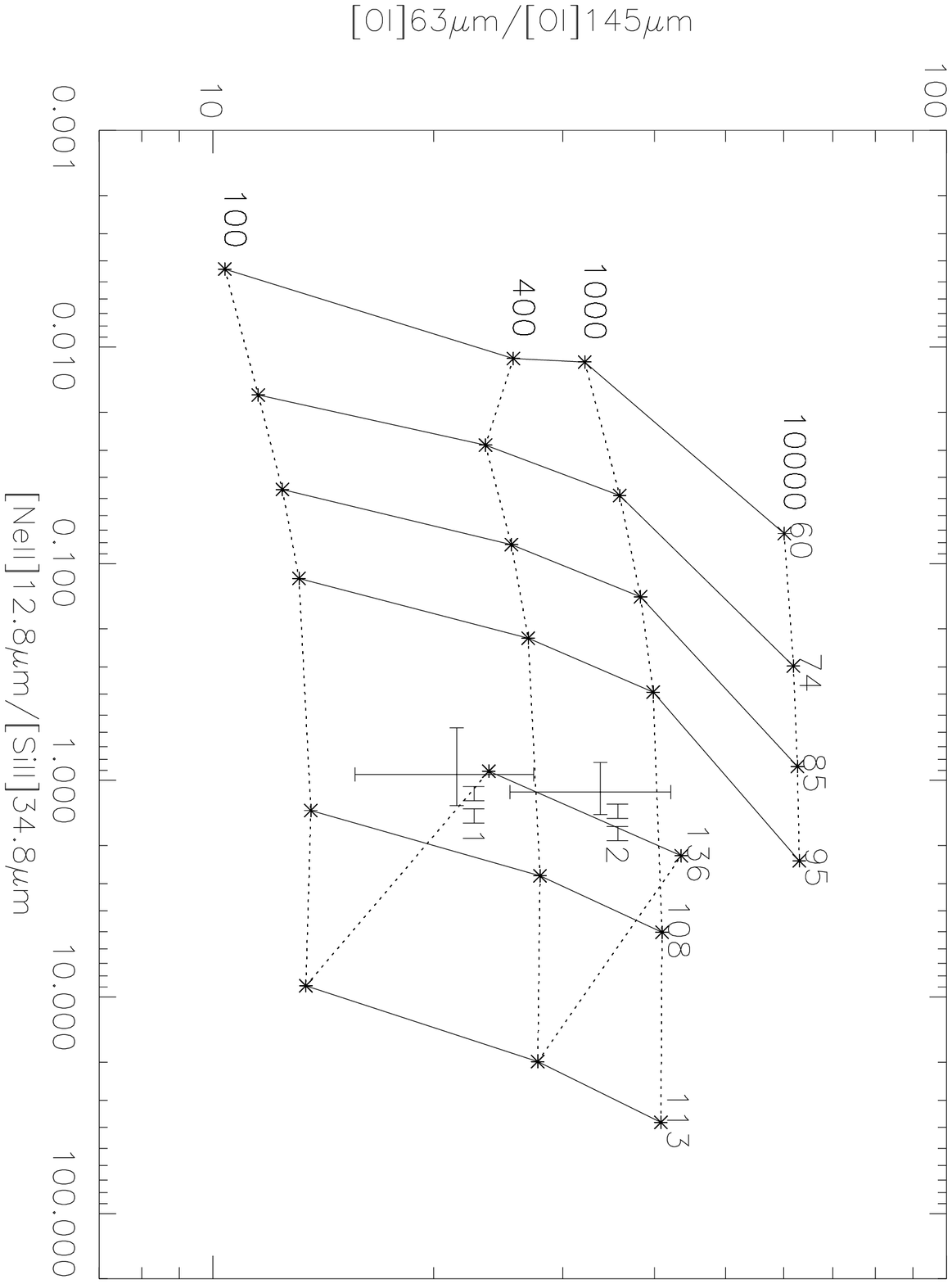]{\label{hh1-2_diag} Diagnostic diagram showing the model results for the indicate
line ratios; full lines are the loci of constant shock velocity (from 60 to 136 \kms), while dashed lines are
loci of constant pre-shock density (from 10$^2$ to 10$^4$ \cmthree).}

\figcaption[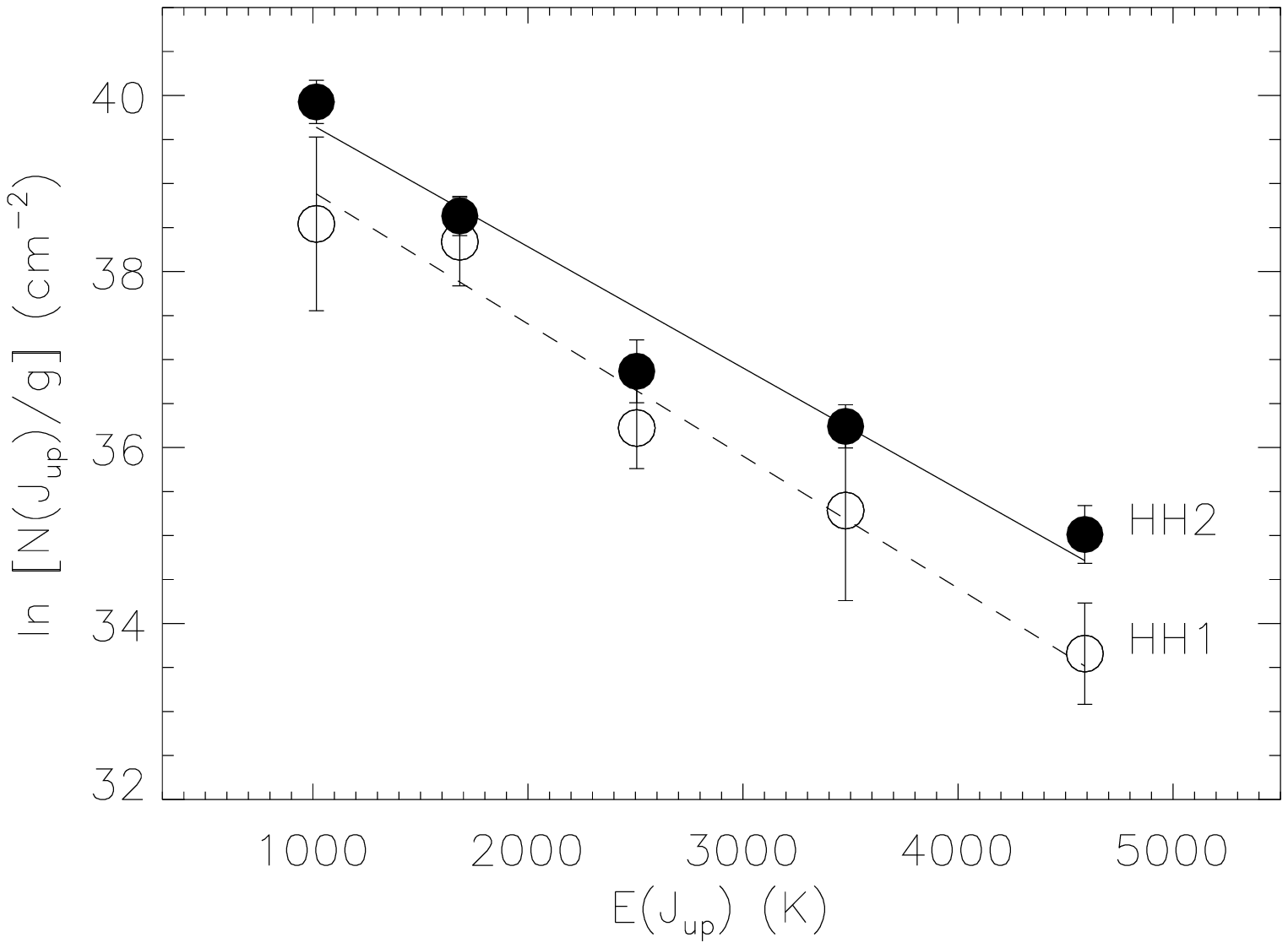]{\label{h2_excitation} H$_2$ excitation diagrams (Boltzmann plots) for HH\,1 and
HH\,2 (empty and full circles respectively). $N_j$/$g_j$ is reported as a function of the energy of the level in
K. The full lines represent linear regressions to the data points.}

\clearpage
%
%

%

%
%
\begin{deluxetable}{lrccc}
\tablecaption{Observed Line Fluxes\label{restab}}
\tablewidth{0pt}\tablehead{
\colhead{Line} & \colhead{$\lambda$(\um)} & \colhead{VLA\,1} &
\colhead{HH\,1} & \colhead{HH\,2}}
\startdata
\cutinhead{SWS Lines}
(0-0)S(1)    & 17.03 & \nodata & 0.7(0.3) & 2.8(0.3) \\*
(0-0)S(2)    & 12.28 & \nodata & 2.3(0.5) & 3.1(0.3) \\*
(0-0)S(3)    & 9.66  & \nodata & 3.0(0.6) & 5.8(0.9) \\*
(0-0)S(4)    & 8.03  & \nodata & 1.8(0.8) & 4.7(0.5) \\*
(0-0)S(5)    & 6.91  & \nodata & 3.6(0.9) & 14(2)    \\*

[Ne{\sc ii}] & 12.81 & \nodata & 4.7(0.9) & 6.8(0.7) \\*

[Si{\sc ii}] & 34.82 & \nodata & 5(1)     & 6(1)     \\*
\cutinhead{LWS Lines}
CO 18-17     & 144.8 & 2(1)     &          &          \\*
CO 17-16     & 153.3 &          &          & 3.0(0.6) \\*
CO 16-15     & 162.8 & 6.8(0.7) & 5(2)     & 4(2)     \\*
         & & \it{(5.7)}  & \it{(4)}   & \it{(3)} \\*
OH[3/2,5/2-3/2,3/2] & 119.3 & & & 4(1) \\*
OH[3/2,1/2-1/2,1/2] & 163.2 & & & 3(2) \\*

o-H$_2$O[2$_{12}$-1$_{01}$] & 179.5 & & & 7(2) \\*

[O{\sc i}] & 63.18 & 173(2) & 101(2) & 150(3) \\*
         & & \it{(151)}  & \it{(86)}   & \it{(135)} \\*

[O{\sc i}] & 145.5 & 5(1) & 4(1) & 4(1) \\*
         & & \it{(4)}  & \it{(4)}  & \it{(4)}  \\*

[C{\sc ii}] & 157.7 & 79(1) & 90(3) & 82(3) \\*
            & & \it{(63)}  & \it{(84)}  & \it{(76)}  \\*
\enddata
\tablecomments{Units of 10$^{-20}$ \lflux and 1$\sigma$ uncertainties in parenthesis.}
\end{deluxetable}

\begin{deluxetable}{lcc}
\tablecaption{H$_2$ Physical Parameters\label{h2physparam}}
\tablewidth{0pt}\tablehead{
\colhead{Object} & \colhead{T$_{\rm H_2}$} & \colhead{N(H$_2$)} \\
\colhead{~} & \colhead{(K)} & \colhead{(10$^{19}$ cm$^{-2}$)}}
\startdata
HH\,1    & 670(40) & 1.1(0.4) \\
HH\,2    & 730(20) & 2.2(0.3) \\
\enddata
\tablecomments{Based on aperture-corrected line fluxes (see text).}
\end{deluxetable}

%
%
\clearpage

\begin{figure}
\plotone{Molinari.fig1.ps}
\end{figure}
\begin{figure}
\plotone{Molinari.fig2a.ps}
\end{figure}
\begin{figure}
\plotone{Molinari.fig2b.ps}
\end{figure}
\begin{figure}
\plotone{Molinari.fig2c.ps}
\end{figure}
\begin{figure}
\plotone{Molinari.fig3.ps}
\end{figure}
\begin{figure}
\plotone{Molinari.fig4.ps}
\end{figure}
\begin{figure}
\plotone{Molinari.fig5.ps}
\end{figure}
\begin{figure}
\plotone{Molinari.fig6.ps}
\end{figure}

\end{document}